\documentclass[12pt,a4paper]{article}

\usepackage{epsfig}
\usepackage{amssymb}
\usepackage{array}
\usepackage{amsmath}
\usepackage{graphicx}

\newcommand{\ncmd}{\newcommand}

\newtheorem{defi}{Definition}
\newtheorem{theo}{Theorem}
\newtheorem{prop}{Proposition}
\newtheorem{lem}{Lemma}
\newtheorem{cor}{Corollary}

\newtheorem{rem}{Remark}

\newcommand{\SO}[1]{\mbox{$\mathbf{SO}(#1)$}}

\newcommand{\bea}{\begin{eqnarray}}
\newcommand{\eea}{\end{eqnarray}}
\newcommand{\nn}{\nonumber}
\ncmd{\btheo}{\begin{theo}$\!\!\!$. -- }
\ncmd{\etheo}{\end{theo}}
\ncmd{\bpro}{\begin{prop}$\!\!\!$. -- }
\ncmd{\epro}{\end{prop}}
\ncmd{\preuve}{{\sc Preuve --}\ }
\ncmd{\bdefi}{\begin{defi} $\!\!\!$. -- }
\ncmd{\edefi}{\end{defi}}
\ncmd{\bco}{\begin{cor}$\!\!\!$. -- }
\ncmd{\eco}{\end{cor}}
\ncmd{\ble}{\begin{lem}$\!\!\!$. -- }
\ncmd{\ele}{\end{lem}}
\ncmd{\bno}{\begin{notation}$\!\!\!\!\!$. -- }
\ncmd{\eno}{\end{notation}}
\ncmd{\bre}{\begin{rem}$\!\!\!$. --  \begin{em}}
\ncmd{\ere}{\end{em} \end{rem}}
\ncmd{\beq}{\begin{equation}}
\ncmd{\eeq}{\end{equation}}
\ncmd{\ben}{\begin{enumerate}}
\ncmd{\een}{\end{enumerate}}
\ncmd{\bit}{\begin{itemize}}
\ncmd{\eit}{\end{itemize}}
\ncmd{\refp}[1]{(\ref{#1})}

\ncmd{\Fi}{\mathbb{F}}
\ncmd{\Oc}{\mathbb{O}}
\ncmd{\Ha}{\mathbb{H}}
\ncmd{\R}{\mathbb{R}}
\ncmd{\C}{\mathbb{C}}
\ncmd{\Z}{\mathbb{Z}}
\ncmd{\N}{\mathbb{N}}
\ncmd{\Sph}{\mathbb{S}}
\ncmd{\T}{\mathbb{T}}
\ncmd{\D}{\mathbb{D}}
\ncmd{\Lp}{\mathfrak{p}}
\ncmd{\Lg}{\mathfrak{g}}
\ncmd{\La}{\mathfrak{a}}
\ncmd{\Lk}{\mathfrak{k}}
\ncmd{\Lm}{\mathfrak{m}}
\ncmd{\Lh}{\mathfrak{h}}
\ncmd{\tr}{\mbox{tr}}
\ncmd{\ad}{\mbox{ad}}
\ncmd{\Ad}{\mbox{Ad}}
\ncmd{\diff}{\mbox{Diff}(M)}
\ncmd{\End}{\mbox{End}}
\ncmd{\Exp}{\mbox{Exp}}
\ncmd{\HH}{\mbox{H}}
\ncmd{\V}{\mbox{V}}
\ncmd{\riem}{\mbox{Riem}(M)}
\ncmd{\Aut}{\mbox{Aut}}
\ncmd{\Id}{\mbox{Id}}
\ncmd{\I}{\mathcal{I}}
\ncmd{\Ker}{\mbox{Ker}}
\ncmd{\di}{\displaystyle}
\ncmd{\bs}{\backslash}
\ncmd{\ov}{\overline}
\ncmd{\no}{\noindent}
\ncmd{\ra}{\rightarrow}
\ncmd{\lra}{\longrightarrow}
\ncmd{\eps}{\epsilon}
\ncmd{\M}{\mathcal{M}}
\ncmd{\DD}{\mathcal{D}(M)}
\ncmd{\super}{\mathcal{S}}

\ncmd{\scalar}[2]{\mbox{$\mathcal{h} #1,#2 \mathcal{i}$}}
\ncmd{\ichap}{\^{\i}}

\title{A Note on the Topology of a Generic Subspace of Riem}
\author{Henrique de A. Gomes\footnote{University of Nottingham, School of Mathematical Sciences, gomes.ha@gmail.com, pmxhg3@nottingham.ac.uk
}}

\begin{document}
\maketitle

\begin{abstract}
For $\M:=\riem$  the space of Riemannian metrics over a compact 3-manifold without boundary $M$,  we study
topological properties of the dense open subspace $\M'$ of metrics which possess no Killing vectors. Given the
stratification of $\M$, we work under the condition that, in a sense defined in the text, the connected components
of each stratum do not accumulate.
 Given this condition we
find that one of the most fundamental results regarding the topology of $\M$, namely that it has trivial homotopy
groups,
   would still be true for $\M'$. This would make the
  topology of $\M'$ completely understood. Coupled with the fact that
  for $\M'$, $\diff\hookrightarrow \M' \overset{\pi}{\ra}\M'/\diff$ is a principal fiber bundle, which makes
  $\M'/\diff$ a proper manifold (as opposed to $\M/\diff$), we would have that $\pi_n(\M'/\DD)=\pi_{n-1}(\DD)$,
  which  reflects the topology of $M$.
  These results would render the space of metrics with no symmetries subject to the above condition,
  $\M'$, as an ideal setting for
  geometrodynamical analysis.
\end{abstract}

\section{Introduction}
Geometrodynamics, as championed by Wheeler, is the study of gravitation through a primary focus on {\it space and
changes therein} rather than on space-time itself. It is in essence merely a dynamical view of GR,
 technically taking form as its constrained Hamiltonian formulation.

To make such an analysis tractable (and for its numerous nice  causality properties) space-time is assumed to be
globally hyperbolic and
 hence to be homeomorphic to a topological product $M\times\R$, with
$M$ being a space-like hypersurface. As such the (unconstrained) configuration space is given by
$$\M:=\riem=~~~\mbox{the space of all 3-Riemannian metrics over}~~ M~$$
In the main body of this paper this will be the space we will work with; the space  of all possible descriptions
of spatial configurations of the vacuum Universe.

Defining
 $$\DD:=\diff~~~\mbox{the space of all 3-diffemorphisms of}~~ M~$$  we identify in $\M$
the 3-metrics\footnote{We reserve the $g$'s to the three metric in this paper since it will be the only type that
appears throughout the text.} $h=f^*g$ and $g$, where $f\in \DD$ acts by pull-back. The resulting orbit space
$$\super:=\M/\DD ~~~\mbox{is called superspace, the space of geometries of $M$}$$
This purports to be the physical space, and projection leaves physical quantities invariant under spatial
diffeomorphisms.

 However a problem quickly arises; the space of spatial geometries, $\super$, is not
properly a manifold \cite{Fischer}, which makes the use of many structures inconvenient. However, we will  very
briefly describe earlier work showing that $\super$ is a stratified manifold; where the strata correspond to
manifolds of metric indexed by the (conjugacy classes of) their symmetry groups.

Thus the following is properly a manifold \cite{Ebin}: $\M'/\DD=\super'$, where $\M'$ is a generic subspace (and a
proper infinite-dimensional manifold) of the space of metrics consisting only of the metrics which possess no
non-trivial symmetry group:
$$\M':=\{g\in\M~|~I_g(M)=\Id\}~~~~\mbox{space of  metrics without non-trivial isometries.}
$$ Then, it is furthermore possible to construct a principal fiber bundle
structure
$$\DD\hookrightarrow \M' \overset{\pi}{\ra}\M'/\DD=:\super'$$
which makes many tools from gauge theory available for the study of the dynamics of $\super'$.

 The  subspace $\M'$ of $\M$ has not been greatly explored in the literature (see \cite{Freed} for an exception),
  in spite of these
 interesting properties. One of the reasons might be because $\M$ has a very simple topology, namely it is
 contractible, a property which might not have extended to any non-trivial subspace. Another alternative has been
 to use the resolution of the singularities of stratification through the use of the augmented space $\M\times
 F(M)$ \cite{Fischer1986} to study the topology of  $\M$, and its relation to the topology of $M$ \cite{Giulini:1993ui}.

 Here we prove that if the condition stated in the abstract
 \footnote{Namely, whenever the number of connected components of each
strata is infinite, no infinite sequence of points
 belonging to different connected components of the strata have an accumulation point in the same strata.}, which we call condition $\mathcal{C}$ in the text,
  holds true,
 then this contractibility property of $\M$, which  is extremely
 valuable in all kinds of analysis on it,
   extends to $\M'$.

   More specifically, using the genericity of $\M'$, the fact that
  $\M$ is contractible, the slice theorem for $\M$ \cite{Ebin}, and using an infinite-dimensional
  Banach version of Thom's transversality theorem,
    we are able to prove that under condition $\mathcal{C}$ $n$-spheres in $\M'$ are contractible.
     Furthermore, since for $\M'$, unlike $\M$,
    we have a proper
    principal fibration, we get the interesting result that
    topological properties of $\super'$ depend only on topological
properties of $\DD$.

In section 2 we give the basic setting and some well-known theorems in the study of $\M$ which will be relevant to
our main theorem. In section 3, after briefly introducing the main tool of our theorem (which is not very present
in the study of $\M$), Thom's transversality theorem, we proceed to the proof.
\section{Basic Setting}
In this section we introduce the necessary background necessary for our main result.

First of all, some notation. As mentioned in the introduction, $M$ will be a 3-dimensional connected, compact
manifold without boundary. We list some definitions along with some basic properties  of the spaces we have to
deal with in order to investigate the gauge structure of Riem: \bit
\item $M$ is a compact oriented manifold without boundary.
\item $L^2_S(TM):=TM^*\otimes_STM^*$
 and $L^{2+}_S(TM)$ its positive-definite subspace.
\item $\Gamma^r(L_S^{2}(TM))$ , $0<r<\infty$, is a Banach space, separable in the $C^r$-weak (Whitney) topology
(uniform convergence up to $r$-derivatives).\footnote{It can be given the structure of a Hilbert space, with
derivatives up to order $r$ defined almost everywhere and each partial derivative being square integrable.} We
define the Banach manifold $\M^r=\Gamma^r(L_S^{2+}(TM))$ in the same way.
\item  $\bigcap_{r=0}^\infty\Gamma^r(L_S^{2}(TM))=:\Gamma^\infty(L_S^{2}(TM)):=S_2(M)$
is a  Frech\'et space  (Metrizable Complete Locally Convex Topological Vector space), constructed from the inverse
limit of separable Banach manifolds (ILB).
 $\M:=\riem=\bigcap_{r=0}^\infty\M^r$ is an open cone in $S_2(M)$ in the sense that for $\lambda_i>0$ and $g_i\in\M$, $\sum_ig_i\in\M$.
  \eit

\subsection{Orbits}

Let us also review the following general facts, which characterize the action of what will play the role of a Lie
algebra and Lie group \cite{Ebin}:
\begin{itemize}
\item The set $\DD:=\diff$ of smooth diffeomorphisms of $M$ is a regular Lie group,  and it acts on $\M$
 on the right as a group of transformations by pulling back metrics:
\begin{eqnarray*}
\mu:\M\times \DD &\ra& \M \\
~(g,f)&\mapsto & f^*g
\end{eqnarray*}
an action which is smooth with respect to the $C^\infty$-structures of  $\M$ and $\DD$\footnote{The natural
action is on the right since of course $(f_1f_2)^*g=f_2^*f_1^*g$.}. It is clear that two metrics are isometric
if and only if they lie in the same orbit, $$g_1\sim g_2\Leftrightarrow g_1,g_2\in \mathcal{O}_g:=\mu_g(\DD)$$
\item The derivative of the orbit map
$\mu_g: \DD \ra \M$ at the identity \bea
T_{\Id}\mu_g: \Gamma(TM) &\ra& T_g\M \nn\\
\label{jmath}X&\mapsto & L_Xg \eea where $X$ is the infinitesimal generator of a given curve of
diffeomorphisms of $M$.    The spaces $V_g$, tangent to the orbits will be called vertical and are defined as:
$$ V_g:=T_g(\mathcal{O}_g)=\{L_Xg~|~X\in \Gamma(TM)\}
$$ Since $M$ is compact, every $X\in\Gamma(TM) $ is complete and
$\Gamma(TM)$ forms a regular Lie algebra under the usual commutator of vector fields,
$[X_1,X_2]\in\Gamma(TM)$.
\item
For an isometry group $I_gM\in\DD$, the Lie algebra $T_{\Id}I_gM$ is a finite dimensional (and hence
splitting) subspace of $\Gamma(TM)$, and so $$T_{\Id}\mu_g: \Gamma(TM)/T_{\Id}I_gM \ra V_g ~~\mbox{ is an
isomorphism}$$ Furthermore, not only is  $\mu_g: \DD/I_gM \ra \M $ an injective immersion onto
$\mathcal{O}_g$, but more importantly, {\it it is also a homeomorphism} \cite{Ebin}. This is extremely
important as it does not allow different strata to wind around infinitely close to each other, a fact which is
needed for the slice theorem and which we will use for our result on the topology of $\M'$.
\end{itemize}
 So, for example for $g\in \M'$  the
map $ T_{\Id}\mu_g:\Gamma(TM) \ra V_g$ is an isomorphism over the image.

We state one of the most fundamental results regarding the geometry of $\M$, which is needed to show $\M'$ is open
dense:
\begin{theo}[Slice for $\super$,
\cite{Ebin}]\label{slice} For each $g\in\M$ there exists a contractible submanifold $S$ of $\M$ containing $g$
such that
\begin{enumerate}
\item $f\in I_gM\Rightarrow f^*S=S$
\item $f\notin I_gM\Rightarrow f^*S\cap S=\emptyset$
\item There exists a local cross section $\tau:Q\subset \DD/I_g(M)\ra\DD$ where $Q$ is an open neighborhood of the
identity, such that the following map is a
diffeomorphism \bea F: Q\times S &\ra& \mathcal{U}_g\\
([f],s)&\mapsto& \tau([f])^*s \eea where $\mathcal{U}_g$ is an open neighborhood of $g\in\M$.
\end{enumerate}\end{theo}
A straightforward corollary of this theorem is the following:
\begin{cor}\label{cor}
Given $g\in \M$, and an arbitrary neighborhood ${V}$ of $\Id\in\DD$, there exists a neighborhood
$\mathcal{N}_{{V}}$ of $g$ such that for any $h\in \mathcal{N}_{{V}}$, there exists $f\in {V}$ such that
$$ f^{-1}(I_hM)f\subseteq I_gM$$
\end{cor}
The proof follows easily by noticing that we can take $\mathcal{N}_{{V}}\subseteq \mathcal{U}_g$ and
$\tau(Q)\subseteq V$ in the theorem above. Then by the third item in the theorem there exists an element $[f]\in
Q$ such that $h=F([f],s)$, hence $h'=(\tau([f]^{-1}))^*h\in S $. Then by the second item, any element of $\DD$
that fixes $h$ must be in $I_gM$, thus $I_{h'}M=f^{-1}(I_hM)f\subseteq I_gM$. This proves that there always exists
a neighborhood of a metric $g$ where all elements have less symmetry than $g$. Clearly this implies that $\M'$ is
open. To prove that it is dense, one must merely infinitesimally perturb the metric to dissolve any symmetries.

\subsection{Stratification Theorem}

  Fischer, in his Stratification Theorem \cite{Fischer}, has shown that
$\super$ is not actually a manifold, since geometries which possess an isometry beyond the identity don't have
neighborhoods homeomorphic to neighborhoods of less symmetric geometries. The metrics that do allow isometries
impede the quotient space $\M/\DD$ to have a manifold structure.

To be more specific, let  $I_g(M)\subset\DD$ be the isometry group of $g$. For a representation
$\rho_{\mbox{\tiny{G}}}:G\ra \DD$, the conjugacy class of this representation denoted by
$(\rho_{\mbox{\tiny{G}}})$ is made up of the representations $\rho'_{\mbox{\tiny{G}}}:G\ra\DD$ such that
$\rho'_{\mbox{\tiny{G}}}=f\circ\rho_{\mbox{\tiny{G}}}\circ f^{-1}$ for some $f\in\DD$. In this way one defines
inequivalent actions of $G$ as given by non-conjugate representations of $G$. We denote the equivalence class of
$rho_{\mbox{\tiny{G}}}$ of conjugate representations by $[\rho_{\mbox{\tiny{G}}}]$.

 Let $\M_{\rho_{\mbox{\tiny{G}}}}:=\{g \in
\M~|~I_g(M)= \rho_{\mbox{\tiny{G}}}(G)\}$ be the space of metrics with isometry $I_g(M)$.  We define
$\M_{[{\rho_{\mbox{\tiny{G}}}}]}:=\{g \in \M~|~I_g(M)= f\circ\rho_{\mbox{\tiny{G}}}(G)\circ f^{-1}~|~f\in\DD\}$,
is the subspace of metrics which have symmetry type $[\rho_{\mbox{\tiny{G}}}]$. The space of metrics that have $G$
as a symmetry group is $\cup_\beta\M_{[\rho_{\mbox{\tiny{G}}}^\beta]}$ where $\rho_{\mbox{\tiny{G}}}^\beta:G\ra
\DD$ are inequivalent representations. The quotient space for each conjugacy class ${[\rho_{\mbox{\tiny{G}}}]}$,
$\M_{[\rho_{\mbox{\tiny{G}}}]}/\DD$ does form a submanifold, called a stratum. Unfortunately, it is modeled over a
different topological vector space for each symmetry group, namely,
\beq\label{dimension}\frac{S_2(M)}{(\Gamma^\infty(TM)/T_{\Id}I_gM)}\eeq

Even if there exists only one equivalence class of actions of $G$, and thus
$\M_G=\M_{[\rho_{\mbox{\tiny{\tiny{G}}}}]}$, we might still have that the number of components of
$\M_{[\rho_{\mbox{\tiny{G}}}]}$ is infinite. For the moment, let us call the connected components of
$\M_{[\rho_{\mbox{\tiny{G}}}]}$ by $C_i(\M_{[\rho_{\mbox{\tiny{G}}}]})$. {\it We will need the assumption that if
the number of connected components is infinite, then no infinite sequence of points $c_i\in
C_i(\M_{[\rho_{\mbox{\tiny{G}}}]})$ belonging to different connected components have an accumulation point in
$\M_{[\rho_{\mbox{\tiny{G}}}]}$}.
\begin{defi}\label{defi}If for every stratum $\M_{[{\rho_{\mbox{\tiny{G}}}}]}$ the number of connected components
$C_i(\M_{[{\rho_{\mbox{\tiny{G}}}}]})$ is finite, or if infinite, then no infinite sequence of points $c_i\in
C_i(\M_{[\rho_{\mbox{\tiny{G}}}]})$ belonging to different connected components have an accumulation point in
$\M_{[\rho_{\mbox{\tiny{G}}}]}$, we will say {\it $\M$ obeys condition $\mathcal{C}$.}
\end{defi}

Using the above corollary of the slice theorem, \ref{cor} stating the hierarchical relation
 between the symmetry of metrics and their
 neighborhoods,
Fischer showed that:
 \beq\label{strata}\M_{[{\rho_{\mbox{\tiny{G}}}}]}\cap\overline{\M_{[{\rho'_{\mbox{\tiny{G}}}}]}}
 \neq\emptyset\Leftrightarrow [{\rho_{\mbox{\tiny{G}}}}]\leq [{\rho'_{\mbox{\tiny{G}}}}]
 (\mbox~or~ [{\rho'_{\mbox{\tiny{G}}}}]\leq
[{\rho_{\mbox{\tiny{G}}}}])\Leftrightarrow
 \M_{[{\rho'_{\mbox{\tiny{G}}}}]}\subset \overline{\M_{[{\rho_{\mbox{\tiny{G}}}}]}}~\mbox{(resp.}~
 \M_{[{\rho_{\mbox{\tiny{G}}}}]}\subset\overline{\M_{[{\rho'_{\mbox{\tiny\tiny{G}}}}]}})\eeq
which is what gives it its stratification status.

One can resolve the singularities in this construction by considering the space $\M\times F(M)$ \ where $F(M)$ is
the bundle of frames over $M$. One then has the PFB $\DD\hookrightarrow\M\times
F(M)\rightarrow\super_{F(M)}=\{[g,b]~|~(g,b)\simeq (f^*g,f^*b)\in \M\times F(M)\}$ \cite{Fischer1986}. The main
thing here is that even though the action of the diffeomorphisms is not free in either component, it is free on
the product, since an isometry that fixes a frame has to be the identity \cite{O'Neill}. Our initial aim for
developing this work was to study the connection associated to $\DD$ and $\M$, which is why we disregard this
auxiliary $\SO3$ construction.

For any $M$, the  metrics that do not admit isometries beyond the identity, i.e. in $\M'$, form a generic, (open
and dense), subset of $\M$ \cite{Ebin}. The same is true of the respective projected subset, $\super'$, consisting
of unsymmetrical geometries of $\super$. Clearly $\M'$ contains all of its orbits. To emphasize, we work with
$\M'$  instead of $\M$ since  $\M'/\DD$ is properly a manifold, and $\M'$ properly a principal fiber bundle above
$\super'$, hence amenable to gauge theory.

\section{Transversal Submanifolds and Contractibility of $\M'$}

Now we develop the related topology of $\M'$, proving that it is in fact simply-connected. In fact, it will prove
trivial to extend  the simple-connectedness result to prove that all homotopy groups are trivial.

\subsection{Transversal Banach Manifolds}
We need first state a weaker version of Thom's transversality theorem \cite{Thom}, valid in finite dimensions.
  Two finite-dimensional submanifolds $P,Q$  of a given smooth finite-dimensional manifold $N$ will be said to
  intersect transversally, $P\pitchfork Q$,  if at every point of intersection their separate
  tangent spaces at that point together generate the tangent space of the ambient manifold at
  that point: $T_xP\oplus T_xQ=T_xN$. Manifolds that do not intersect are vacuously transverse.
\begin{theo}[Thom]\label{Thom}
Let $M$ be a compact finite-dimensional manifold, $N$ be any smooth manifold and $f_0:M\ra N$. If $Z$ is an
embedded submanifold of $N$ then:
\begin{itemize}
\item {\bf Stability:} If $f_0$ is transversal to $Z$, then for any homotopy $f_t$, there exists $\epsilon>0$ such that for $t<\epsilon$ , $f_t$ is still transversal.
\item {\bf{Genericity:}} There exist a homotopy $f_t$ of $f_0$ and an $\epsilon>0$ arbitrarily small such that  $f_\epsilon$ is transversal to $Z$.
\end{itemize}
\end{theo}

Usually,  transversality theory in finite dimensions gives generic and stable properties of transversality between
any two smooth maps from arbitrary manifold domains into a common target manifold (neither one is necessarily an
embedding). Even though  the whole of transversality theory does not convert to the infinite-dimensional Banach
manifold setting\footnote{Unless, that is, we have two extra conditions: i) one of the domains, say the manifold
$Q$, is finite-dimensional, $f:Q\ra\mathcal{N}$, and ii) the second map,  between arbitrary Banach manifolds,
$\mathcal{F}:\mathcal{P}\ra\mathcal{N}$, is a  smooth Fredholm map (which means its tangent map has everywhere a
finite-dimensional kernel and cokernel) \cite{Smale}. For more details on which properties of transversality
theory can be transplanted to Banach manifolds see \cite{Donaldson}, Ch. 4 (Proposition 4.3.10). The proof of the
following theorem \ref{transversal} itself relies on the fact that Fredholm $C^r$ maps are dense on the space of
continuous maps between suitable Banach manifolds.}, we still have an effective version of Theorem \ref{Thom} in
\cite{Eells1968}(Theorem 1E), which will be sufficient for our purposes:
\begin{theo}[Transversality, \cite{Eells1968}]
Let $X$ and $Y$ be smooth manifolds modeled on a separable Hilbert space $E$ and a Banach space $F$, respectively.
Let $B\subset Y$ be a closed direct submanifold (i.e. such that for every $y\in B$, $T_yB$ is a direct summand of
$T_yY$). Then for any smooth map $\phi:X\ra Y$ and any continuous function $\epsilon:X \ra \R(>0)$ there is a
smooth $\epsilon$-approximation $\Psi:X\ra Y$ of $\phi$ which is transversal over $B$. Furthermore if $K\subset X$
is a closed set on which $\phi$ is transversal over $B$, we can also require that
$\Psi_{|K}=\phi_{|K}$.\label{transversal} \end{theo} Clearly the theorem gives genericity of transversality, and
stability can be recovered as in the original version of the theorem (see sections 4A and 4B of \cite{Eells1968}).

\subsection{Contractibility of $\M'$}

 Now we can state the main theorem regarding the topology of $\M'$:

\begin{theo}[Homotopy groups of $\M'$]\label{theo}
Under the condition  ~$\mathcal{C}$, we have that any  smooth loop in $\M'$ is contractible. I.e. the fundamental
group of $\M'$ is trivial. \end{theo}

 First of all, since the above theorem only works for target
 Banach manifolds, and we have in our hands ILB manifolds, we do the usual adaptation  of
 working with the extensions of $C^\infty(E)$ to $\mathcal{B}^r(E)$, which is the completion of the space
 $C^\infty(E)$ under the $C^r$-Whitney topology, and is thus a Banach space.

 In the following then,
 $\M_{[\rho_\alpha]}^r$ will denote the completion of the manifold $\M_{[\rho_\alpha]}$
  under the $C^r$-Whitney
 topology, and is thus a Banach manifold. Here, to simplify notation,
  $\alpha$ parametrizes all the equivalence classes of all the
 groups, and is countable \cite{Fischer}. In other words for $[\rho_\gamma^\beta]$, we have the bijection
  $(\gamma,\beta)\equiv\alpha\in \N$.

 \begin{lem}\label{lemma}Under the condition $\mathcal{C}$, any closed
 compact manifold of $\M^r$ has a finite number of intersections with $\{\overline{\M_{[\rho_\alpha]}^r},\alpha\in
 \N\}$.  \end{lem}
 {\em{Proof of Lemma \ref{lemma}}.}
 First of all a comment is in order about the extendability of the Slice and Stratification theorems.
 {The slice theorem \cite{Ebin} itself is proven first using such completions, i.e. for $\M^r$
 (see Theo. 7.1.), but in that case, completion with respect to an inner product induced by the
 $r$-th jet bundle, thus making it a Hilbert manifold. We do not need this extra structure for the present purpose.
 The Stratification theorem relies basically on the hierarchical structure present in the slice theorem (item iii),
 and is hence also easily extendable to the Banach completion.

  Let
$\Lambda\in \N$ label a particular set of $[\rho_\alpha]$, where $\M^r_{[\rho_\alpha]}$ are the maximal
submanifolds (in the partial ordering of conjugacy classes of Lie groups mentioned above) that intersect a given
compact submanifold $K$ of $\M^r$. We are only interested in the number of intersections between each maximal
submanifold $\overline{\M^r_{[\rho_\alpha]}}$ and $K$, intersections which, by \eqref{strata} are disjoint for
different $[\rho_\alpha]$ and non-accumulating for the same $[\rho_\alpha]$ by assumption. Under this assumption,
there will be no difference in proof if we  assume there is indeed only one connected component of each
$\M^r_{[\rho_\alpha]}$.

 By implicitly using the Axiom of Choice, we form a sequence of unique
representatives from each intersection, $\{g_\alpha~|~g_\alpha\in K\cap\overline{\M^r_{[\rho_\alpha]}}, \alpha\in
I\subset\Lambda\}$. Since $K$ is compact, either the sequence is finite or it has an accumulation point in $K$.
Suppose that there is an accumulation point $g_\infty$.

Then  we have that there  necessarily exist different $g_{\mbox{\tiny N}}\in \overline{\M^r_{[\rho_{\mbox{\tiny
N}}]}}$ and $g_{\mbox{\tiny N'}}\in \overline{\M^r_{[\rho_{\mbox{\tiny N'}}]}}$ which lie in the same open set
domain $U^r_{g_{\mbox{\tiny N}}}$ of a slice at $g_{\mbox{\tiny N}}$. But by  corollary \ref{cor} we then have
that $I_{g_{\mbox{\tiny N'}}}(M)$ and $I_{g_{\mbox{\tiny N}}}(M)$ are conjugate, which would imply an inclusion
relation between $[\rho_{\mbox{\tiny N}}]$ and $[\rho_{\mbox{\tiny N'}}]$. This is absurd, since we specifically
chose the maximal representatives in each tree. If there are more than one connected component of the strata, as
by assumption there is no accumulation point, the number of intersections of each strata with $K$ will also be
finite.

 Thus, under the conditions $\mathcal{C}$,
  {\it any
 compact manifold of $\M^r$ has a finite number of intersections with $\{\overline{\M^r_{[\rho_\alpha]}},\alpha\in \N\}$}.
 ~$\square$\medskip

 {\em{Proof of Theorem \ref{theo}}.}
  Now take $\alpha:S^1\ra {\M^r}'$ to be
an embedding. As ${\M^r}'$ is open $\alpha$ exists. We know that $\M^r$ is a smooth manifold, it is in fact a
contractible cone inside a topological locally convex vector space. Thus there exists an embedding of the disc
$D^2$ given by $f_0:D^2\ra\M^r$ such that $f_0(\partial D^2)=\alpha(S^1)$. We shall abbreviate notation and
sometimes call $f_0$ the surface when in fact we mean the image of $f_0$.

 Let us start by considering the submanifold $\overline{\M^r_{(G_{\alpha_1})}}$, as in the above lemma. We take
 a homotopy of the embedding $f_0$, call it $(f_0)_t$. By genericity of
 transversality\footnote{Note that even though one of the embeddings is finite-dimensional, and the kernel of
 the inclusion
 map of the $M_{[\rho_\alpha]}$ is finite-dimensional, it is not necessarily Fredholm,
  since the cokernel may have infinite-dimension. This is why we resort to the weaker theorem in
  \cite{Eells1968}.
  See the `Remark' in
   \cite{Eells1968}, pg 5. },
there exists an arbitrarily small $\epsilon_1$ such that we have an embedding $(f_0)_{\epsilon_1}$, which is
homotopic to $f_0$ where $(f_0)_{\epsilon_1}$ is everywhere transversal to the closure of $\M^r_{(G_{\alpha_1})}$.
We rename $(f_0)_{\epsilon_1}=f_{\epsilon_1}$. Furthermore, by theorem \ref{transversal}, we can choose it to
satisfy ${f_{\epsilon_1}}_{|\partial D^2}={f_{0}}_{|\partial D^2}$ which maintains the boundary of this embedding
as $\alpha$.

 Thus we have a surface in $\M^r$ whose boundary is $\alpha$ and which is
everywhere transversal to $\overline{\M^r_{[\rho_{\alpha_1}]}}$.

Now by the stability property of transversality, for each homotopy of $f_{\epsilon_1}$,  there exists a `stability
distance' $\epsilon>0$ such that for $t<\epsilon$ it will remain transversal to
$\overline{\M^r_{[\rho_{\alpha_1}]}}$. Then we repeat the first process for $\overline{\M^r_{[\rho_{\alpha_2}]}}$;
for some homotopy $(f_{\epsilon_1})_t$, where by genericity we now find an $\epsilon_2<\epsilon_1'$, such that
$\epsilon_1'$ is within the stability distance for transversality between $f_{\epsilon_1}$ and
$\overline{\M^r_{[\rho_{\alpha_1}]}}$. Hence $(f_{\epsilon_1})_{\epsilon_2}$ is transversal to
$\overline{\M^r_{[\rho_{\alpha_2}]}}$ and we have kept it transversal to $\overline{\M^r_{[\rho_{\alpha_1}]}}$ as
well. Finite iteration will get us a surface, which we call $f_\lambda$, whose boundary is $\alpha$ and which is
everywhere transversal to $\{\overline{\M^r_{[\rho_\alpha]}},\alpha\in\N\}$.

Now we are left to prove that transversality between $\overline{\M^r_{[\rho_\alpha]}}$ and the constructed
two-dimensional submanifold $f_\lambda$ implies that their intersection is vacuous.  Recall that the
$\M^r_{[\rho_\alpha]}$ are splitting submanifolds (at each point $y\in \M^r_{[\rho_\alpha]}$ the entire tangent
space $T_y\M^r$ is a direct summand of its tangent space $T_y \M^r_{[\rho_\alpha]}$), hence we can form the
projection $\pi: T_y\M^r\ra T_y\M^r/T_y \M^r_{[\rho_\alpha]}$.

Now, suppose that such a $y$ is in the transversal surface, i.e. $y= f_\lambda(x)$. From theorem
\ref{transversal}, we have that the composition
$$\pi \circ T_xf_\lambda:T_xD^2\ra T_y\M^r/T_y \M^r_{[\rho_\alpha]}$$ is surjective.
But $T_y\M^r\simeq S_2(M)^r$ and $T_y \M^r_{[\rho_\alpha]}$ has infinite dimensionality and
co-dimensionality.\footnote{ As a matter of fact, one can see that
$f^*\M_{\rho_{\mbox{\tiny{G}}}}=\M_{\rho'_{\mbox{\tiny{G}}}}$ for
$\rho'_{\mbox{\tiny{G}}}=f\circ\rho_{\mbox{\tiny{G}}}\circ f^{-1}$. In fact, we have that
$$T_y
\M^r_{\rho_\alpha}\simeq \Gamma^r(\frac{TM}{T\rho(\Lg)}\otimes_S \frac{TM}{T\rho(\Lg)})$$ where $u,v\in TM$ are
equivalent if $u=v-T_{\mbox{\tiny Id}}\rho(X)$ for $X\in\Lg$. So $T_y \M^r_{[\rho_\alpha]}$ is modelled on a
subspace of $$ {\mu_g}((\DD/I_gM))_*\left(\Gamma^r(\frac{TM}{T\rho(\Lg)}\otimes_S \frac{TM}{T\rho(\Lg)})\right)$$
where one has to quotient out the action of the diffeomorphism group by the isotropy subgroup.}
 As $D^2$ is only two dimensional, the map
cannot be surjective, which means $f_\lambda^{-1}( \M^r_{[\rho_\alpha]})=\emptyset$. Since  $
\M_{[\rho_\alpha]}\subset \M^r_{[\rho_\alpha]} $ this implies the theorem. ~~~$\square$\medskip

Furthermore,  we can repeat the argument for $D^{k}$ and assert that, under  condition $\mathcal{C}$, {\it all
homotopy groups of $\M'$ are trivial}.

\section{Conclusions}
 We have studied the topology of an open dense subset of riemannian metrics on a compact manifold without
 boundary, namely, the subset of metrics with trivial isometry group. For this study, we used intesively
  both the Ebin-Palais
 Slice theorem \cite{Ebin}, which states the existence of a slice for the action of the diffeomorphism group,
 and Fischer's stratification theorem \cite{Fischer1986}, which states that $\M$ is partitioned into
 submanifolds of metrics
 with different symmetries.

If the different connected components of the strata are non-accumulating in the sense of definition \ref{defi}, we
proved that $\M'$ is contractible.
 The strategy of the proof was to
 find discs $B^{n+1}$ with fixed border $S^n$ contained in $\M$, which were transversal to all such submanifolds of
 metrics possessing symmetry. Then finding  that
 transversality meant vacuous intersection is a trivial step.
 To be able to apply the
 transversality theorem we had at hand \cite{Eells1968}, we had to use these theorems in the weaker,
 Banach completion
 domain.

It is known that differential topology of infinite-dimensional separable Hilbert manifolds is completely
classified, in the sense that if $H_1$ and $H_2$ are infinite-dimensional separable Hilbert manifolds and have
isomorphic homotopy groups, they are in fact diffeomorphic. Hence in our case, since all our homotopy groups are
trivial we can in fact state that $\M'$ is itself contractible.

Furthermore, since we now also would have properly a principal fiber bundle $\diff\hookrightarrow \M'
\overset{\pi}{\ra}\M'/\diff$, following the long exact homotopy sequence we would have that
$\pi_n(\M'/\DD)=\pi_{n-1}(\DD)$. I.e.  if $\M$ obeys condition $\mathcal{C}$, $\M'/\DD$ is a classifying space for
$\DD$. The dependence of related invariants on the topology of $M$ are studied in \cite{Fischer:1991pt}. For some
classes of $M$, $\DD$ is contractible as well, hence $\M'\simeq \super'\times\DD$. If this is the case it can have
no Gribov ambiguities, and coordinate choices such as the harmonic ones are smoothly well-defined over $\M'$.

 {\it
The study of  the dependence of condition $\mathcal{C}$ on the topology of $M$ is left for future work.} We also
leave for future work the ascertainment of the question if condition $\mathcal{C}$  is satisfied for all compact
$M$ without boundary, but only for certain symmetry groups $G_\alpha$. In this case, their complement
$\M_{[\rho_G]}$, would have to be open dense subset of $M'$ for the work done here to remain valid.

As a last remark, it is interesting in fact to compare our result to \cite{Giulini:1993ui}. There, Giulini studies
the case of an open manifold, let us call it $\Sigma$ and its one point compactification $\bar\Sigma=\Sigma\cup
\{\infty\}$. By resolving the singularities of stratification using the extended space Riem$(\bar\Sigma)\times
F(M)$, as we mentioned in passing in section 2, one finds a principal fiber bundle Diff$_{\mbox{\tiny
F}}(\bar\Sigma)\hookrightarrow \mbox{Riem}(\bar\Sigma)\times F(M)
\overset{\pi}{\ra}\mbox{Riem}(\bar\Sigma)/\mbox{Diff}_{\mbox{\tiny F}}(\bar\Sigma)$. By using the same techniques
as we did one arrives at $\pi_n(\mbox{Riem}(\bar\Sigma)/\mbox{Diff}_{\mbox{\tiny
F}}(\bar\Sigma))=\pi_{n-1}(\mbox{Diff}_{\mbox{\tiny F}}(\bar\Sigma))$, where $\mbox{Diff}_F(\bar\Sigma)$ are
diffeomorphisms $f$ such that $f(\infty)=f(\infty)$ and that $f_*(\infty)=\mbox{Id}$. So in that case, there is no
restriction to a subset of Riem, but there is a restriction on the group of diffeomorphisms, i.e.
$\mbox{Riem}(\bar\Sigma)/\mbox{Diff}_{\mbox{\tiny F}}(\bar\Sigma)$ is a classifying space for the group
$\mbox{Diff}_F(\bar\Sigma)$. Note that dependence of these homotopy groups on the topology of $\Sigma$ differs
from the dependence of $\pi_{n}(\DD)$ on the topology of $M$.

Furthermore, we mention in passing that by studying the homotopy of $\M'/\DD$ one in principle could distinguish
between homotopy equivalent manifolds $M_1, M_2$ which are not homeomorphic \cite{Giulini:2009np}. See also this
last reference for numerous possible applications of these concepts to quantum gravity and for a literature
review.

 \paragraph{Acknowledgments}
I  want to thank John Barrett for important feedback.


\end{document}